# Hybrid-Fusion Transformer for Multisequence MRI

Jihoon Cho[0000-0002-8619-9481] and Jinah Park[0000-0003-4676-9862]

KAIST, 291 Daehak-ro, Yuseong-gu, Daejeon 34141, Republic of Korea
{zinic, jinahpark}@kaist.ac.kr

**Abstract.** Medical segmentation has grown exponentially through the advent of a fully convolutional network (FCN), and we have now reached a turning point through the success of Transformer. However, the different characteristics of the modality have not been fully integrated into Transformer for medical segmentation. In this work, we propose the novel hybrid fusion Transformer (HFTrans) for multisequence MRI image segmentation. We take advantage of the differences among multimodal MRI sequences and utilize the Transformer layers to integrate the features extracted from each modality as well as the features of the early fused modalities. We validate the effectiveness of our hybrid-fusion method in three-dimensional (3D) medical segmentation. Experiments on two public datasets, BraTS2020 and MRBrainS18, show that the proposed method outperforms previous state-of-the-art methods on the task of brain tumor segmentation and brain structure segmentation.

**Keywords:** Transformer, Multi-modality, 3D Medical image segmentation

## 1 Introduction

Magnetic resonance imaging (MRI) is widely used in the detection, diagnosis, and treatment planning of diseases in the human body, including the brain, spinal cord, prostate, and knee. Depending on the target organ and purpose, there are several types of MRI protocols consisting of many sequences [18]. Each MRI sequence has shown various characteristics, especially the signal of different tissues such as fluid, muscle, and fat. In addition, some sequences represent functional information beyond the anatomical structure [21]. Considering that a valuable feature varies by sequence type, a combination of sequences gives better results than unimodal processing in the presence of diseases [17] and lesion segmentation [4].

In recent years, Convolutional Neural Networks (CNN) have been successful in various computer vision tasks. U-Net [16] adopts the concept of a FCN [14] with a relatively shallow structure and balancing feature representation and locality through the skip connection. Following the modifications suitable for 3D medical image segmentation, U-net has become the de facto standard even for multimodal MRI volumes with the early fusion of simple multichannel input [9, 10] as shown in Fig. 1 (a). More recently, Vision Transformer (ViT) [3], inspired by the tremendous success of Transformer [19], become a new solution to the limited receptive field of CNN with the global self-attention mechanism. For 3D medical segmentation, UNETR [6] and Swin-UNETR [5] have proposed Transformer networks with CNN layers on the decoder, and TransBTS [20] have constructed with CNN encoder and decoder with the bottom Transformer layer. However, all of these Transformer networks have the disadvantage of treating multiple MRIs as a multichannel input.

Similar to multisequence MRI, RGB-D images consist of multiple modalities that have the same spatial information: color image and depth image. However, considering the sharable and specific features between color and depth images [7], particular encoding (as shown in Fig. 1 (b)) is used for each modality in many tasks including semantic segmentation [11]. This approach has been used even for recent work of Transformer-based methods [12, 13]. From the effort to consider the multimodalities of RGB-D images, we find that the adoption of the middle fusion approach for MRI sequences can benefit from different modality characteristics.

In this work, inspired by the processing of multimodal RGB-D images and the long-range visual dependence from ViT, we propose the Hybrid-Fusion Transformer (HFTrans) for multisequence MRI images. The proposed HFTrans is constructed with the hybrid fusion approach to take advantage of both early fusion and middle fusion, as shown in Fig. 1 (c), and consists of multiple CNN encoders and the Transformer encoder. Each encoder extracts a local context feature representation for each modality, including the early fused modalities, and they are integrated in the Transformer encoder. The feature embedding from the Transformer encoder is progressively up-sampled with the spatial information from encoders via skip-connection, and finally predicts segmentation maps of the original resolution. In experiments on the Brain Tumor Segmentation 2020 dataset (BraTS2020) [2] and the MR Brain Segmentation 2018 dataset (MRBrainS18)[1], we validate the effectiveness of our method in multisequence MRI segmentation. HFTrans achieves remarkable performance on both public challenge datasets. We also conduct further experiments on encoder compositions, which show that our hybrid fusion method works well without human heuristics by using simple encoders for each multisequence MRI image.

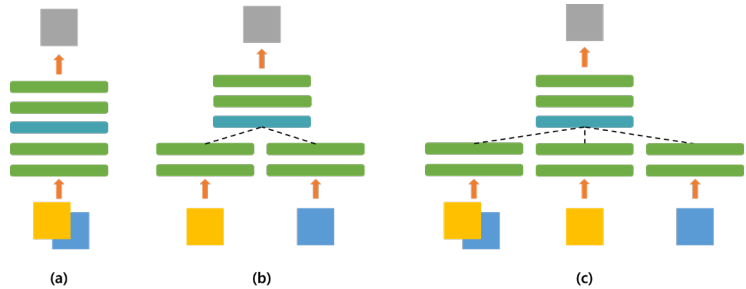

**Fig. 1.** Three types of fusion methods. (a) early fusion (b) middle fusion (c) our hybrid fusion

---

[1] https://mrbrains18.isi.uu.nl/

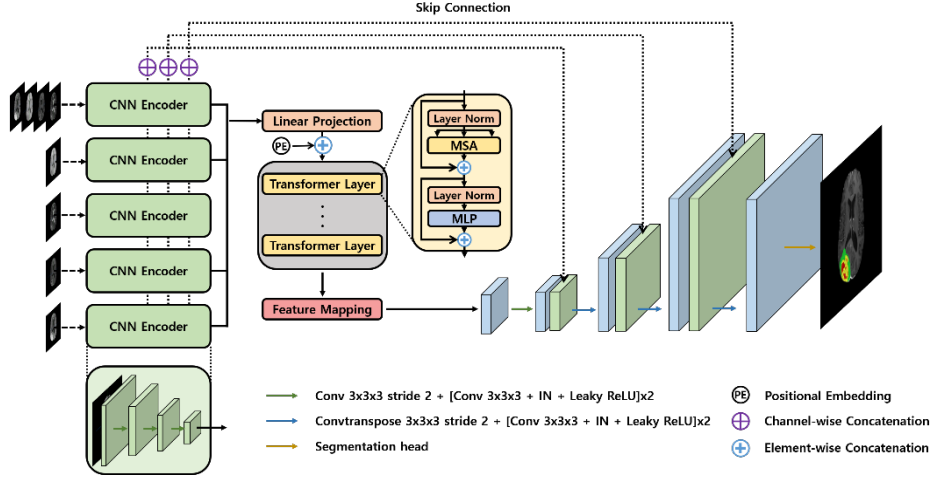

**Fig. 2.** Overview of HFTrans network for BraTS2020 dataset. Hybrid fusion of four MRI sequences is performed in the CNN Encoders and the Transformer Encoder. The encoded representation output from the Transformer encoder is progressively upsampled with skip connection to predict the final segmentation maps.

## 2  Method

An overview of HFTrans is presented in Fig. 2. Although we accept the early fusion encoding, the hybrid fusion method is applied by constructing additional encoders for each modality.

### 2.1  Hybrid Fusion from CNN Encoders

Considering the high computational cost of Transformer for high-resolution 3D images and the inductive bias of the convolutional layer, we propose to construct the convolutional layers to make a rich local context feature representation. To bring benefits from different modality characteristics, each modality is processed in individual encoders. Features are embedded into 1D sequences and then perform self-attention between feature embedding in the Transformer layer. In addition, the encoding of early fused modalities is also utilized, taking into account the ability to extract an apparent powerful representation. Hybrid fusion between powerful representation from the entire encoding and modality-specific representation from separate encoding can exploit the advantages of both methods.

For the 3D MRI input consisting of $N$ MRI sequences $x_i (i = 1, 2, ..., N) \in \mathbb{R}^{1 \times W \times H \times D}$ with resolution (W, H, D), we use $N + 1$ feature representations for hybrid fusion. $N$ features are extracted individually from each MRI sequence, and the representation of early fusion is encoded by all $N$ sequences $x \in \mathbb{R}^{N \times W \times H \times D}$. The encoders have the same structure consisting of stacking the convolutional layers 3×3×3 and stride-convolutional layers consecutively. Then, the high-level feature representations $f_j(i = 1, 2, ..., N + 1) \in R^{K \times \frac{W}{8} \times \frac{H}{8} \times \frac{D}{8}}$ are projected linearly, but the computational complexity of the Transformer layer is increased quadratic based on the number of 1D sequences. Therefore, we apply the 2x2x2 patch embedding projection to the features extracted from CNN. Subsequently, we get the input embedding $z_0 \in R^{C \times M (=N \times \frac{W}{16} \times \frac{H}{16} \times \frac{D}{16})}$ with the channel dimension C. To preserve location information of flattened sequences, we add a learnable positional embedding $E_{pos} \in R^{C \times M}$ as

$$z_0 = W \times f + E_{pos} \tag{1}$$

where W is the 1D projector with 2x2x2 patch. After the feature embedding, we conduct self-attention using a standard Transformer encoder consisting of L Transformer layers. The l-th Transformer layer is operated as follows,

$$z_l^* = MSA(LN(z_{l-1})) + z_{l-1} \tag{2}$$

$$z_l = MLP(LN(z_l^*)) + z_l^* \tag{3}$$

where MSA denotes multihead self-attention, MLP is multilayer perceptron, and LN refers to layer normalization.

## 2.2   CNN Decoder and Loss Function

The output sequences of the Transformer encoder are reshaped in the 4D feature maps to generate voxel-wise semantic segmentation results. The reshaped feature maps $d_j \in R^{C \times \frac{W}{16} \times \frac{H}{16} \times \frac{D}{16}}$ are concatenated channel-wise and upsampled by a factor of 2 to shape the original feature size of fj before linear projection. After feature mapping, the feature representation $d \in R^{K \times \frac{W}{8} \times \frac{H}{8} \times \frac{D}{8}}$ is progressively fed into the deconvolution of stride 2 and the convolutional layers 3x3x3. During deconvolution, we aggregate the encoding features of multiple CNN encoders via a skip connection. This process is repeated up to the feature representation reaching the original input resolution, and the final semantic segmentation is generated through the convolutional layer 1x1x1 with a softmax activation function. We use both Dice loss and cross-entropy loss together as an objective function.

## 3      Experiments

### 3.1    Datasets

We use two publicly available 3D medical segmentation datasets consisting of multi-modal MRI images: BraTS2020 and MRBrainS18.

**BraTS2020**: BraTS2020 [2] is a patient's brain MRI dataset labeled with three tumor sub-regions, peritumoral edematous tissue, enhancing tumor, and necrotic tumor core. The dataset contains 369 training sets acquired from several institutions with various protocols and scanners. Each MRI scan consists of four sequences: T1-weighted (T1), T2-weighted (T2), post-contrast T1-weighted (T1ce), and T2 fluid-attenuated inversion recovery (FLAIR). They were provided after preprocessing of the co-registration and skull stripping, and we additionally perform z-score normalization to brain regions except the masked background area with zero intensity. All MRI sequences have the same voxel size of 240×240×155 with 1mm isotropic voxel spacing.

**MRBrainS18**: For the whole brain segmentation, we use the MRBrainS18 dataset that includes both brain structure and pathological abnormalities. The dataset consists of 30 subjects acquired on a 3T scanner from various patients, including dementia, diabetes, and Alzheimer's. Multimodal MRI scans consist of aligned sequences of T1, T1 inversion recovery sequence (T1-IR), and FLAIR. All scans have a 0.958mm× 0.958mm×3mm voxel spacing with 240×240×48 voxel size. We perform a 7-fold cross-validation for 7 training set and use 8 labels in the evaluation, which are gray matter, basal ganglia, white matter, white matter lesion, CSF, ventricles, cerebellum, and brain stem.

Table 1. Cross-validation results on the BraTS2020 dataset. ET, TC, and WT denote enhancing tumor, tumor core, and whole tumor.

| Models | Dice Score (%) ↑ | | | | HD95 (mm) ↓ | | | |
|---|---|---|---|---|---|---|---|---|
| | ET | TC | WT | Avg. | ET | TC | WT | Avg. |
| U-Net [16] | 80.79 | 80.40 | 88.67 | 83.29 | 32.20 | 17.13 | 4.15 | 17.83 |
| ResUNet [22] | 80.31 | 78.45 | 88.79 | 82.52 | 29.05 | 16.11 | 4.70 | 16.62 |
| AttnUNet [15] | 80.73 | 79.30 | 88.54 | 82.86 | 31.24 | 23.86 | 11.30 | 22.13 |
| nnU-Net [8] | 82.28 | 84.18 | 90.56 | 85.67 | 32.20 | **5.03** | 2.68 | 13.30 |
| TransBTS [20] | 81.39 | 80.70 | 90.16 | 84.08 | 30.59 | 15.17 | 7.68 | 17.81 |
| HFTrans | **82.81** | **84.66** | **90.82** | **86.10** | **26.42** | 6.98 | **2.57** | **11.99** |
| HFTrans* | 82.52 | 84.59 | 90.40 | 85.84 | 29.79 | 5.61 | 3.98 | 13.13 |

## 3.2  Quantitative Results

We perform experiments on BraTS2020 and MRBrainS18 datasets by comparing our HFTrans with five previous state-of-the-art: (1) U-Net [16]; (2) ResUNet [22]; (3) AttnUNet [15]; (4) nnU-Net [8]; (5) TransBTS [20], which is the Transformer-based network with an early fusion approach. We perform a five-fold cross-validation on the BraTS2020 dataset for all methods. As shown in Table 1, HFTrans achieves Dice scores of 82.81%, 84.66%, 90.82% and HD95 of 26.42mm, 6.98mm, 2.57mm on ET, TC, WT, which are higher results than the other methods except HD95 of TC. Compared to U-Net [16], ResUNet [22], AttnUNet [15], TransBTS [20], and nnU-Net [8], our proposed method outperforms them by 2.81%, 3.58%, 3.24%, 2.02% and 0.53% in terms of average Dice score and 5.84mm, 4.63mm, 10.14mm, 5.82mm, and 1.31mm in terms of average HD95, respectively. HFTrans*, the hybrid fusion variant model that consists of modality exception encoders instead of each modality encoder (described in Table 3), also outperforms the previous methods.

The results evaluated on MRBrainS18 are reported in Table 2. HFTrans achieves Dice score 84.81%, HD95 3.25mm, and volume similarity 94.12%, which outperforms the result of nnU-Net [8] and TransBTS [20] by 2.16% and 1.44% in terms of Dice score, 3.29mm and 2.01mm in terms of HD95, and 1.49% and 1.08% in terms of volume similarity. It is also comparable to U-Net [16], ResUNet [22], and AttnUNet [15]. Comparing the model complexity, U-Net, ResUNet, AttnUNet, and our HFTrans have 90.30M, 37.72M, and 25.78M, 65.17M parameters and 266.91G, 498.53G, 329.54G, and 140.39G FLOPs, respectively. Despite the relatively small model complexity, HFTrans shows significantly better performance, especially in brain stem segmentation, by bridging high-level global context information with low-level local details.

**Table 2.** Cross-validation results on the MRBrainS18 dataset. Note: GM: gray matter, BG: basal ganglia, WM: white matter, WML: white matter lesions, CSF: cerebrospinal fluid, Vent: ventricles, Cereb: cerebellum, BS: brain stem.

| Models | Dice Score (%) ↑ | | | | | | | | |
|---|---|---|---|---|---|---|---|---|---|
| | GM | BG | WM | WML | CSF | Vent | Cereb | BS | Avg. |
| U-Net [16] | **84.79** | 83.89 | 86.24 | **64.86** | 82.82 | 93.60 | **92.62** | 88.59 | 84.67 |
| ResUNet [22] | 84.23 | 83.51 | 86.06 | 64.27 | 82.29 | 93.29 | 92.17 | 88.88 | 84.34 |
| AttnUNet [15] | 84.68 | 83.17 | 86.51 | 63.44 | 82.54 | **93.69** | 92.46 | 89.50 | 84.50 |
| nnU-Net [8] | 82.60 | 80.99 | 85.49 | 60.59 | 79.87 | 92.59 | 91.00 | 88.09 | 82.65 |
| TransBTS [20] | 83.07 | 83.71 | 85.78 | 60.59 | 80.43 | 92.57 | 92.35 | 88.47 | 83.37 |
| HFTrans | 84.71 | 83.74 | **86.99** | 64.03 | 82.35 | 93.50 | 92.32 | **90.85** | **84.81** |
| HFTrans* | 84.33 | **84.17** | 86.80 | 63.80 | 82.46 | 93.59 | 91.40 | 90.60 | 84.64 |

| Models | HD95 (mm) ↓ | | | | | | | | |
|---|---|---|---|---|---|---|---|---|---|
| | GM | BG | WM | WML | CSF | Vent | Cereb | BS | Avg. |
| U-Net [16] | **0.96** | 3.07 | **1.15** | 10.83 | **1.98** | 1.36 | 1.36 | 3.85 | 3.30 |
| ResUNet [22] | 1.01 | 3.10 | 1.51 | 10.76 | 2.04 | 1.53 | 1.53 | 3.95 | 3.37 |
| AttnUNet [15] | **0.96** | 3.11 | 1.48 | 10.95 | **1.98** | 1.48 | 1.48 | 3.46 | 3.31 |
| nnU-Net [8] | 1.52 | 3.86 | 1.96 | 12.38 | 2.44 | 1.78 | 1.78 | 25.26 | 6.54 |
| TransBTS [20] | 1.18 | 3.02 | 1.84 | 12.60 | 2.41 | 2.12 | 2.12 | 16.02 | 5.26 |
| HFTrans | 1.15 | **2.85** | 1.48 | **10.36** | 2.08 | 2.41 | **2.79** | **2.90** | **3.25** |
| HFTrans* | 1.07 | 3.05 | 1.47 | 11.15 | **1.98** | 1.36 | 3.23 | 3.21 | 3.32 |

| Models | Volume Similarity (%) ↑ | | | | | | | | |
|---|---|---|---|---|---|---|---|---|---|
| | GM | BG | WM | WML | CSF | Vent | Cereb | BS | Avg. |
| U-Net [16] | 95.16 | 94.91 | 94.42 | **82.96** | 94.95 | 96.91 | 96.24 | 93.63 | 93.49 |
| ResUNet [22] | 95.20 | 94.06 | 94.43 | 80.61 | 93.53 | 94.44 | 95.91 | 95.04 | 93.28 |
| AttnUNet [15] | 95.48 | 93.68 | 95.29 | 79.39 | **95.09** | 96.63 | 95.96 | 94.27 | 93.30 |
| nnU-Net [8] | 95.37 | 93.47 | **96.12** | 76.55 | 94.38 | 97.46 | 95.19 | 93.01 | 92.63 |
| TransBTS [20] | 95.73 | 94.81 | 96.02 | 73.17 | 95.03 | **97.53** | **97.49** | 94.67 | 93.04 |
| HFTrans | **95.74** | 94.16 | 96.04 | 80.59 | 94.97 | 96.92 | 96.63 | 96.03 | **94.12** |
| HFTrans* | 95.17 | **95.49** | 95.61 | 78.30 | 94.53 | 97.14 | 95.38 | **96.13** | 93.55 |

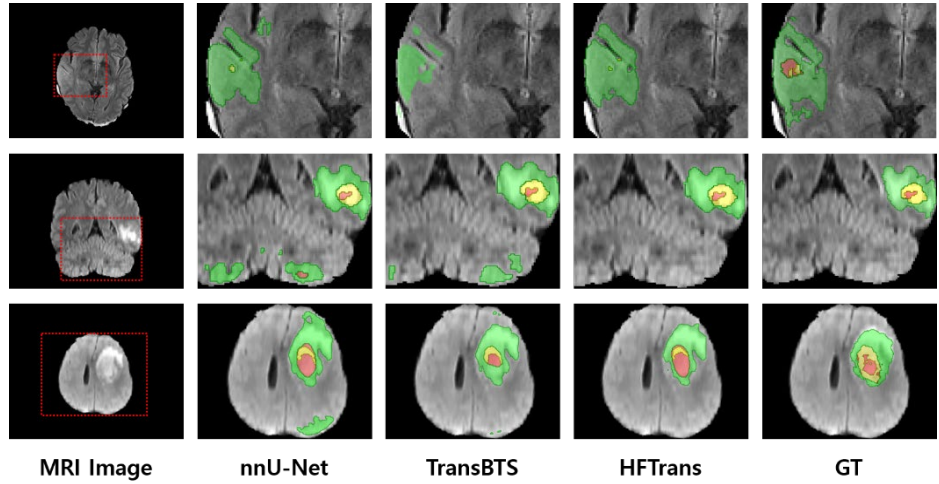

**Fig. 3.** Qualitative comparison of brain tumor segmentation on the BraTS2020 dataset. The enhancing tumor (ET) is depicted in the yellow region, and the tumor core (TC) is represented as a union of red and yellow regions. The whole tumor (WT) contains a colored region of green, red, and yellow.

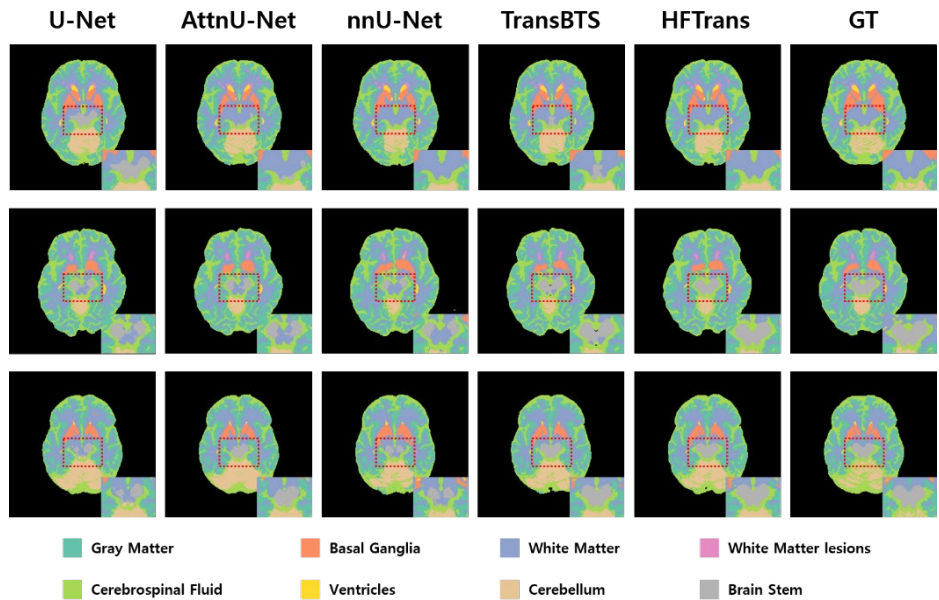

**Fig. 4.** Qualitative comparison of brain structure segmentation on the MRBrainS18 dataset. The brain stem region (gray) is zoomed-in.

### 3.3    Qualitative Results

Qualitative comparisons on brain tumor segmentation are presented in Fig. 3. Our hybrid fusion method HFTrans shows fine-grained segmentation of brain tumors, while the pure CNN-based method nnU-Net tends to over-segment and the CNN-Transformer method TransBTS tends to under-segment, which are evident in rows 1 and 3. This indicates that hybrid fusion captures both powerful spatial context and long-range dependency. In Fig. 4, we present qualitative segmentation comparisons for brain structure segmentation in the MRBrainS18 dataset. HFTrans exhibits detailed segmentation of the whole brain structure. In particular, our method shows superior performance with a detailed boundary in brain stem segmentation, and the effectiveness of the hybrid fusion method is demonstrated.

**Table 3.** Results for different variants of encoder composition. We compare the early fusion method, the middle fusion method, and our hybrid fusion methods including the additional human heuristics. T1* denotes the three-channel input of T2, T1ce, and FLAIR except for T1. T2*, T1ce*, and FLAIR* have the same approach as T1*.

| Encoder Composition | Dice (%) | HD95 (mm) |
|---|---|---|
| T1 / T2 / T1ce / FLAIR (Middle Fusion) | 82.40 | 28.01 |
| All (Early Fusion) | 83.06 | 25.56 |
| All / T1ce | 82.62 | 27.35 |
| All / FLAIR | 83.02 | 24.09 |
| All / T1ce / FLAIR | 83.17 | 25.62 |
| All / T1+T1ce / FLAIR | 82.58 | 27.29 |
| All / T1 / T2 / T1ce / FLAIR (HFTrans) | **83.52** | 24.07 |
| All / T1* / T2* / T1ce* / FLAIR* (HFTrans*) | 83.28 | **22.63** |

## 4    Discussion

We evaluate the effectiveness of our encoder composition by comparing the early fusion approach, which takes all modality as input, the middle fusion approach of individual feature extraction from modalities, and considering the human heuristics of the annotation protocol [1], that the appearance of a brain tumor is typically depicted as a hyperintense signal in T1ce and FLAIR. As shown in Table 3, the middle fusion approach shows the worst results of Dice score 82.40% and HD95 28.01mm, failing to get the benefit of each modality. The early fusion approach shows the better results of 83.06% and 25.56mm in terms of Dice score and HD95. Several results of different human heuristic approaches improve HD95 of 3.92mm when using the early fusion encoder and additional FLARE encoder, and improve Dice score of 0.77% when using the early fusion encoder, FLARE and T1ce encoders. However, they do not produce an improvement for both the Dice score and HD95 at the same time compared to the early fusion approach. The encoder compositions of our proposed method HFTrans, taking advantage of early fusion and middle fusion, improve performance by 1.12% and 3.94mm on Dice Score and HD95 without human heuristics. In addition, the variant of our method, HFTrans*, also shows improvements in both metrics, especially with a remarkable HD95 result of 22.63mm.

## 5    Conclusion

This paper introduces a novel Transformer-based segmentation framework for multisequence MRI. The proposed hybrid fusion method inherits the advantages of the early fusion approach with the powerful locality of 3D CNN and the middle fusion approach with the global consistency of Transformer. Experiments on different volumetric segmentation datasets, BraTS2020 and MRBrainS18, validate the effectiveness of our method. The proposed method could serve as the basis for a Transformer-based segmentation network for multimodal medical images. As a future work, we plan to explore the Transformer-based fusion method with a focus on the computational efficiency.

**Acknowledgement.** This work was supported by Korea Institute of Energy Technology Evaluation and Planning(KETEP) grant funded by the Korea government(MOTIE) (20201510300280, Development of a remote dismantling training system with force-torque responding virtual nuclear power plant).